\documentclass[10pt]{article}

\usepackage[colorlinks,urlcolor=blue,linkcolor=blue,citecolor=blue]{hyperref}
\expandafter\def\expandafter\UrlBreaks\expandafter{\UrlBreaks\do\/\do\*\do\-\do\~\do\'\do\"\do\-}

\usepackage[utf8]{inputenc}
\usepackage{algorithm}
\usepackage{algpseudocode}
\usepackage{graphicx}
\usepackage{amsfonts}
\usepackage{amssymb}
\usepackage{hyperref}
\usepackage{authblk}
\usepackage{booktabs}

\begin{document}

\title{Effort and Size Estimation in Software Projects with Large Language Model-based Intelligent Interfaces}

\author[1,2]{Claudionor N. Coelho Jr}
\author[1]{Hanchen Xiong}
\author[1]{Tushar Karayil}
\author[1]{Sree Koratala}
\author[1]{Rex Shang}
\author[1]{Jacob Bollinger}
\author[1]{Mohamed Shabar}
\author[1]{Syam Nair}
\affil[1]{Zscaler Inc.}
\affil[2]{ECE Department, Santa Clara University}

\date{\today}
\maketitle

\begin{abstract}
The advancement of Large Language Models (LLM) has also resulted in an equivalent proliferation in its applications. Software design, being one,  has gained tremendous benefits in using LLMs as an interface component that extends fixed user stories. However, inclusion of LLM-based AI agents in software design often poses unexpected challenges, especially in  the estimation of development efforts. Through the example of UI-based user stories, we provide a comparison against traditional methods and propose a new way to enhance specifications of natural language-based questions that allows for the estimation of development effort by taking into account  data sources, interfaces and algorithms. 
\end{abstract}

\section{INTRODUCTION}
The acceleration of LLM model development and their visibility have prompted the genesis of many LLM-based products. Recently, the release of ChatGPT~\cite{chatgpt2022,GPT4TechnicalReport} was a milestone that signaled a significant shift in society, including changes in software design paradigms. Initially, LLMs~\cite{Huyen2023} like ChatGPT revolutionized the field with advanced chatbots 
and AI Agents~\cite{wolfram2023}, enhancing the ability of these models by connecting data sources, algorithms and visualizations to LLMs. However, in recent months many have witnessed a transition towards more sophisticated systems such as Retrieval-Augmented Generation (RAG)~\cite{lewis2021retrievalaugmented} 
and AI Agents~\cite{wolfram2023}.

Although more recent LLMs~\cite{openai2023gpt4,touvron2023llama,jiang2024mixtral} have the capability to do data analysis and even data summarization and representation, the ability to connect to external data sources, algorithms and specialized interfaces to LLMs~\cite{nexusflow2023} adds additional flexibility to LLMs by enabling it to perform tasks that involves analysis of domain specific real time data, or even the possibility to perform tasks that are still beyond LLM's capabilities.

This paper discusses the changes in software design using AI Agents, specifically, the shift from traditional UI/UX user stories~\cite{userstories:2004} in software design to LLM-based AI Agent interfaces implementing several user stories using a single natural language interface. This transition represents a paradigm shift from well-structured documentation of data sources, UI/UX interactions, and algorithms, where you can reasonably well estimate size and effort of development, to a more flexible, albeit imprecise, mode of interaction through natural language descriptions. While this shift has unlocked unprecedented levels of user accessibility and software adaptability, it has also introduced unique challenges.  One of the most fundamental questions that we intend to address in this paper is on how to estimate the development effort and size of these new systems, where the LLM interacts with the user sometimes in unknown ways.

\section{UI/UX BASED SYSTEM DESIGN AND EFFORT ESTIMATION}

In this section we provide a simple example to show how effort can be estimated using current software engineering methods~\cite{sommerville, wppf, voorhees2021guide}.  We do not intend to show how to compute efforts, but we only emphasize here that knowing the number of data sources, user interface widgets and algorithms enables one to estimate the effort and size of a project or feature.

\begin{figure}[htpb]
    \centering
    \includegraphics[width=4.5cm]{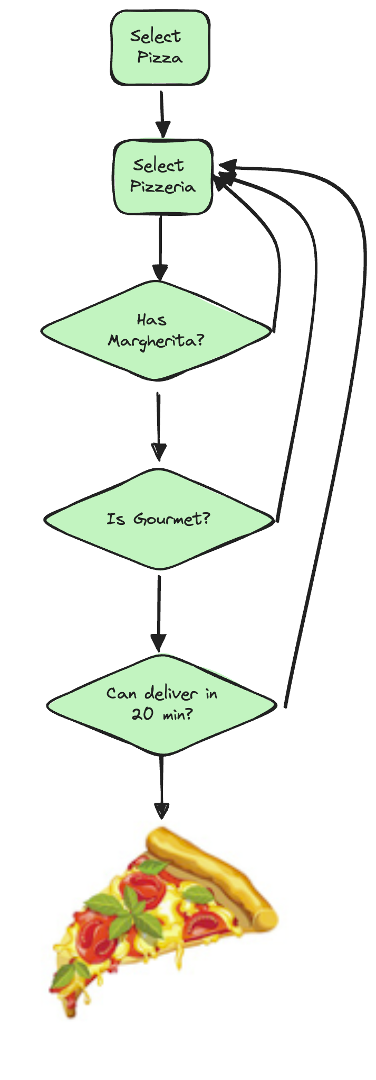}
    \caption[User Story]{User Story to Order Pizza for a Food Delivery App} 
    \label{fig:order-pizza}
\end{figure}

In this example, we want to examine the complexity of adding the user story of ordering a margherita gourmet pizza in 20 minutes to a food app, as 
an optimization to the the flow presented in Figure~\ref{fig:order-pizza}.

We have to assume that to implement this use case, we need access to the following data sources and algorithms:

\begin{enumerate}
    \item Restaurant database that can be searched by location and by type of food.
    \item Menu database, where user can search for types of food served by the restaurant.
    \item Algorithm that computes the delivery time from the restaurant to your location.
\end{enumerate}

Based on this information, and the number of widgets available in the user interface, we can estimate the development effort based on previous experiences from the team as mentioned in Table~\ref{tab:effort-metrics}.  In this example, we are not considering other data sources, UI widgets algorithms, but in reality, they would be required, such as payment infrastructure.

\begin{table}
\vspace*{4pt}
\label{tab:effort-metrics}
    \centering
    \begin{tabular}{ccc}
\toprule 
        Tables &  Algorithms & Widgets \\ \hline
        \\ 2 &  1 & 4 
    \end{tabular}
\caption{Summary of Effort Metrics}
\end{table}

The reader should notice that this use case implements a single type of user interaction, and if we decide to modify the interaction, we will need to change the user story, or create another implementation that accommodates a different user story.

\section{AI AGENTS}

An AI Agent~\cite{nvidia2023} encompasses a system that employs an LLM to process and reason about a specific domain. To generate specific answers (often related to the domain), the AI Agent leverages auxiliary systems in conjunction with the LLM. These auxiliary systems support the agent in comprehending the domain and facilitating the creation of accurate responses.

AI Agents consists of four major components. The \emph{agent core} 
forms the central component and is responsible for orchestrating the agent's overall functionality. The \emph{memory} module enables the agent to store and retrieve relevant information, enhancing its ability to retain context and make informed decisions. The \emph{planner} component guides the agent's actions by formulating a strategic course of actions based on the given problem or task. Finally, the set of \emph{tools} encompasses various external components and resources that assist the agent in performing specific tasks or functions within the defined domain. These components collaboratively enable AI Agents to effectively process information, reason, and generate responses in a manner aligned with their designated purpose.

\begin{figure}[htp]
    \centering
    \includegraphics[width=7cm]{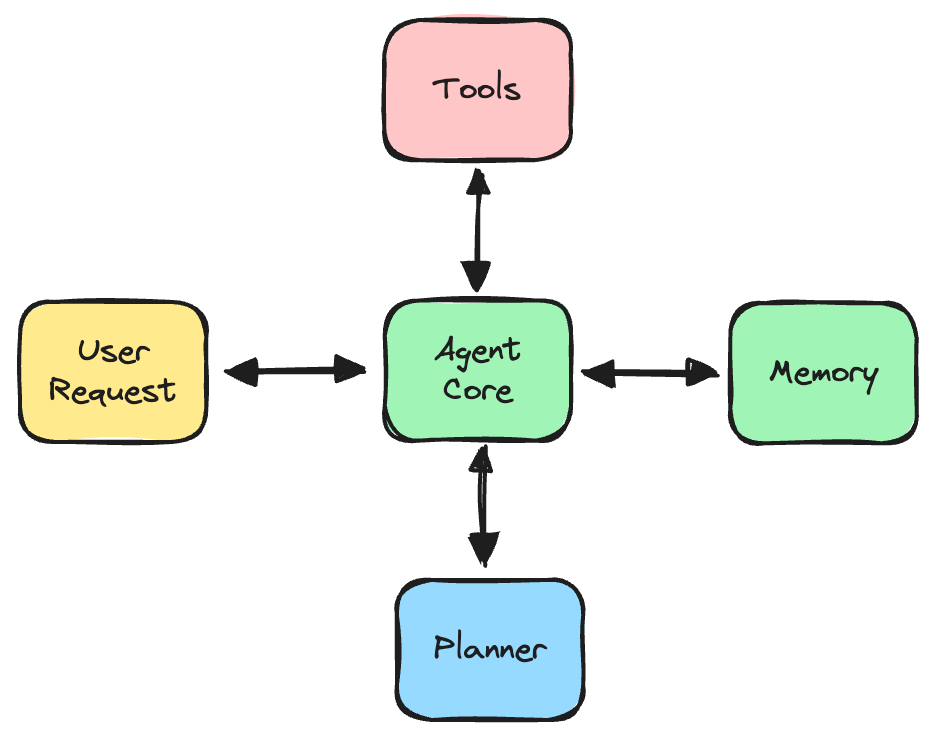}
    \caption{AI Agent from~\cite{nvidia2023}}
    \label{fig:ai-agent}
\end{figure}

\subsection*{Agent Core}

The agent core is a crucial component within an AI Agent that plays a central role in orchestrating the agent's overall functionality. It receives a query from the user. Consequently, it manages the decision-making processes, communication, and coordination of various modules and subsystems within the agent. Finally, it aggregates the information and generates a response. 


The agent core is also responsible for managing the agent's internal state. It maintains a representation of the agent's assets and internal state, allowing it to reason, plan, and adapt its behavior accordingly. The core oversees the update and retrieval of information from the agent's memory, enabling it to access relevant knowledge and contextual information during decision-making processes.


\subsection*{Memory}

The memory module within an AI Agent encompasses two important aspects: historical memory and contextual memory. 

{\bf Historical memory} serves as a repository for past interactions and experiences of the AI Agent. It stores a record of previous inputs, outputs, and the outcomes of actions taken by the agent. This historical data is valuable as it enables the agent to learn from past interactions and avoid repeating mistakes. Through the historical memory the agent gains insights about effective strategies, successful outcomes/patterns enabling an informed decision making process. 

{\bf Contextual memory}, on the other hand, focuses on maintaining a coherent understanding of the current situation. It stores relevant context that provides the necessary background for the agent to interpret and respond appropriately to the present state. This can include information about the environment, the user's preferences or intentions, and any other contextual factors that influence the agent's behavior. Contextual memory allows the agent to adapt its action and responses to specific circumstances, thereby enhancing its ability to interact intelligently with changing environments.

Together, historical and contextual memories allow the AI agent to combine past experiences and current context for an efficient decision making process



\subsection*{Planner}

The planner component within an AI Agent plays a crucial role in guiding the agent's actions and formulating a strategic course of action based on the given problem or task. It is responsible for generating a sequence of steps or actions that lead the agent towards achieving its objectives. The planner analyzes the current state of the environment, along with any available information or constraints, to determine the most effective sequence of actions to achieve the desired outcome. It also takes into account other factors such as goals, resources, rules, and dependencies to generate a plan that optimizes the agent's decision-making process.

An example of a prompt template that can be used by the planner is presented in Figure~\ref{fig:prompt}.  Please note that we use the \emph{Model-View-Controller} (MVC) architecture~\cite{voorhees2021guide} as a convenient way to describe data, interfaces and algorithms, respectively, as LLMs have probably been exposed to this framework during training phase. The planner would then utilize this prompt template to generate a plan that outlines specific actions and steps to be taken. By employing the planner component, the AI Agent can systematically determine the optimal sequence of actions to achieve its objectives, ensuring efficient decision-making and effective utilization of available resources. The generated plan serves as a roadmap for the agent's actions, enabling it to navigate complex problem spaces and accomplish its goals in an optimal manner.

%

\begin{figure}[htp]
\begin{verbatim}
GENERAL INSTRUCTIONS
You are an autonomous AI Agent who converts a text into 
executable tasks using as few interactions as possible
with the user. Your task is to break down each complex 
request from a list of user requests into simpler tasks. Each
simpler task should use an existing tool or if none is 
avaiable, you should create a helper task. Each task should
be one of model, view or control from software MVC 
architecture. Model or data sources are objects representing
database tables. Models can be searched and modified by 
algorithms. Control are algorithms that represents actions on
data sources. Views or interfaces describe interaction with 
user. They need to take as input either a data source, or an 
algorithm, when some computation needs to be performed on one
or more data sources. Your list of tasks should concisely 
represent the algorithms, data sources and interfaces that 
need to be implemented to perform the task. Your answer 
should be only a csv list with fields task type, function
call name and task description from MVC model and nothing more.


AVAILABLE TOOLS:
- Search Tool
- Math Tool
 
CONTEXTUAL INFORMATION:
<information from Memory to help LLM
to figure out the context>

USER REQUEST:
I want to order margherita pizza in 20 min in my app?
 
ANSWER FORMAT
csv list
\end{verbatim}
    \caption{Example of Prompt by Planner, modified from~\cite{nvidia2023}}
    \label{fig:prompt}
\end{figure}


\subsection*{Tools}

In an AI Agent, the set of tools encompasses various resources and functionalities that assist in performing specific tasks or functions within the defined domain. Here is a non-exhaustive list of possible tools that can be utilized in an AI Agent:

\begin{itemize}
\item {\it RAG (Retrieval-Augmented Generation)} ---  Combines retrieval-based methods with generative language models. It enables the agent to retrieve relevant information from a knowledge base and utilize it to generate coherent and contextually appropriate responses.  Common data sources for RAG include Question-Answer databases, documentation and web pages.
\item {\it Database connections} --- Connect to databases and allow the AI Agent to access and retrieve information from structured external data sources. This tool enables the agent to query and extract relevant data for decision-making or generating responses, specially in domain specific scenarios.
\item {\it Machine Learning frameworks} --- Provide tools and algorithms for training and deploying machine learning models. These frameworks enable the agent to leverage various machine learning techniques, including supervised learning, unsupervised learning, or reinforcement learning, to enhance its capabilities.
\item {\it Visualization tools} --- Assist in representing and interpreting data or model outputs in a visual format. These tools can help the agent understand complex patterns, relationships, or trends in the data, aiding in decision-making and analysis.
\item {\it Simulation environments} --- Provide a controlled virtual environment where the AI Agent can interact and learn without impacting the real world. These tools allow the agent to practice and refine its skills, test different strategies, and evaluate the potential outcomes of its actions.
\item {\it Data preprocessing tools} --- Help in cleaning, transforming, and preparing raw data before feeding it into the AI Agent. These tools may include techniques for data cleaning, normalization, feature selection, or dimensionality reduction, ensuring the quality and relevance of data used by the agent.
\end{itemize}

These tools enhance LLMs by providing it with specialized functionalities for specific domains.  
It should be noted that these tools can be classified as data sources, visualization artifacts and algorithms.

\section{LLM IS THE NEW UI/UX}

With advent of LLMs in the previous year, we have seen people specifying user stories using natural language, as mentioned before, in the following way:

\begin{verbatim}
    I want to order a gourmet Margherita pizza in 20 minutes.
\end{verbatim}

In user story development, as follow-up questions one would need to document in the development process, we would like to determine.

\begin{itemize}
    \item Which data sources should we connect to?
    \item Which algorithms we need to invoke to solve this request?
    \item Which interfaces are required to implement this user story?
    \item Which other questions we want to be able to solve?
\end{itemize}

We have seen a deterioration of specification quality in user stories when people over abuse the capabilities of adaptability of LLMs and we will show how we can easily lose control of this simple requirement by just slightly changing the question.

\begin{enumerate}
    \item Can this restaurant deliver food in 20 min?
    \item Give me the list of all restaurants that deliver gourmet pizza in 20 min.
    \item Give me the 20 top evaluated restaurants that can deliver gourmet pizza in 20 minutes.
\end{enumerate}

The reader can easily see that the first question requires just a simple yes/no answer. The second question requires a summarization or visualization agent to provide the answer. The third query will require getting data from possibly an additional table for the backend.  Without fully specifying what are the problem the system is trying to solve, and resorting to just a single question (as people expect the LLMs to extrapolate automatically on these questions), estimating the development effort may become an almost impossible task.

\section{ESTIMATING EFFORT IN AI AGENTS}

The main idea of this paper is to show that we can retrieve a similar level of understanding of implementation effort of the user stories if we use the {\bf Planner} of an AI Agent to enumerate the data sources and algorithms we need to use by sampling questions we want to be able to answer with these systems. 

The idea is presented below by iterating over generation of related questions and asking planner to generate sub-tasks for the generated set of questions,

\begin{algorithm}
\begin{algorithmic}
\Require List of questions $Q$
\State ${\it AllTasks} \gets \emptyset$
\For{$q \in Q$}
\State Generate $N$ related questions $Q_q$ from $q$.
\State $T_q \gets {\it Planner}($
\State $\qquad \{ q \} \cup Q_q, {\it current\_tools}=AllTasks, $
\State $\qquad minimize=True)$
\For{$t \in T_q$}
\State $t['task'] \gets {\it data\ source}| {\it algorithm} | {\it UI\  widget}$
\EndFor
\State Manually validate the set $T_q$.
\State ${\it AllTasks} \gets {\it AllTasks} \cup T_q$
\EndFor
\State Manually validate final set of {\it AllTasks}
\end{algorithmic}
\end{algorithm}

Once we iterate over sample questions and extract related questions, we should be able to converge on the set of data sources, algorithms and interface items that are required. As an additional piece of information, we will be able to document what the system will do and not do.  For example, by documenting which data sources we are accessing, we should be able to document explicitly which data sources we will not be accessing in this feature implementation.

Please note that at each step of this procedure, we need users to evaluate the questions generated automatically and the tools required to process the questions, as we may have duplicates, unnecessary tools and hallucinations.

We used the following prompt to generate similar questions to the original question specified as an important user story.

\begin{figure}[htp]
\begin{verbatim}
GENERAL INSTRUCTIONS
You are simulating what a user would want to request in a food 
app. You will generate 6 related questions to the user request. 
Each generated question should cover different aspects of what 
a user may want to do in a real scenario. Each aspect should be 
associated to different data sources, algorithms and interfaces
that needs to be implemented by the app. Your generated questions
should be simple sentences. You should refrain from repeating the 
same contents in different questions. Your answer contain the 
list of generated questions and nothing more.Your answer should 
not contain enumerations or itemized lists.

USER REQUEST
I want to order a gourmet Margherita pizza in 20 minutes.
\end{verbatim}
    \caption{Example of Prompt to generate similar questions}
    \label{fig:prompt-similar-queries}
\end{figure}



This prompt generated the following similar questions in a LLM that the system may be required to process. The reader should note that some of these questions may require use of additional data sources, or execution of different algorithms, or even requiring additional visualization widgets.

\begin{enumerate}
\item Can you provide a list of nearby pizzerias that offer gourmet Margherita pizzas with an estimated delivery time of 20 minutes or less?
\item Are there any ongoing promotions or discounts for a Margherita gourmet pizza available for quick delivery?
\item What are the options for customizing a Margherita gourmet pizza, such as crust type or cheese options, while still ensuring a 20-minute delivery?
\item Can you recommend the top-rated restaurant for a gourmet Margherita pizza based on user reviews and delivery speed?
\item Are there any minimum order requirements or additional fees associated with ordering a single Margherita gourmet pizza for quick delivery?
\item Can you filter the restaurant search?
\end{enumerate}

Table~\ref{fig:raw-list} presents the raw list of tools (data sources, algorithms and user interface items) that were generated from the algorithm outlined before, based on the prompt of Figure~\ref{fig:prompt} enhanced with all the questions with the additional instruction to minimize redundant tasks or tools. It is worth noting that by carefully choosing the Planner, we will be able to get a much better and curated list of tools.

You can see that by just using this procedure, we have been able to document the effort to develop this system using 22 algorithms, 11 data sources, and 11 user interfaces, which includes 
one more user interface for the LLM-based AI Agent.  

\section{CONCLUSIONS}

Over the past year, we have seen an explosion in  the integration of Large Language Models to existing systems (or even the creation of new systems where one of the UI/UX widgets is a natural language interface).

Such systems posed a challenge in normal software engineering practices of effort and size estimation, as the system are not as well documented as it used to be when specifying user stories explicitly. These new specifications started using sample questions on what the system should do, which inherits the ambiguity of written language. It becomes impossible to quantify effort or size of development of such systems, or even to document what the system does and does not do.

We have shown that by using an LLM to generate a list of similar questions, and leveraging the planner state of the AI Agent to create a list of non-duplicated sub-tasks, we are able to regain the same level of precision that user stories and use cases had achieved previously.

\section{ACKNOWLEDGMENTS}
Portions of this document used GPT4 to improve readability, and to automatically generate related questions and algorithms, data sources and UI interfaces.

\def\refname{REFERENCES}

\bibliography{./references.bib}

\begin{thebibliography}{10}

\bibitem{chatgpt2022}
``Introducing chatgpt.'' \url{https://openai.com/blog/chatgpt}, November 2022.
\newblock Accessed on December 19, 2023.

\bibitem{GPT4TechnicalReport}
{OpenAI}, ``{GPT-4 Technical Report}.'' arxiv:2303.08774 [cs.CL], 2023.
\newblock Submitted on March 15, 2023 (v1), last revised on March 27, 2023 (v3); Accessed on May 4, 2023.

\bibitem{Huyen2023}
C.~Huyen, ``Building llm applications for production.'' \url{https://huyenchip.com/2023/04/11/llm-engineering.html}, April 2023.
\newblock Accessed on May 4, 2023.

\bibitem{wolfram2023}
``Wolfram|alpha as the way to bring computational knowledge superpowers to chatgpt.'' \url{https://writings.stephenwolfram.com/2023/01/wolframalpha-as-the-way-to-bring-computational-knowledge-superpowers-to-chatgpt/}, January 2023.
\newblock Accessed on December 19, 2023.

\bibitem{lewis2021retrievalaugmented}
P.~Lewis, E.~Perez, A.~Piktus, F.~Petroni, V.~Karpukhin, N.~Goyal, H.~Küttler, M.~Lewis, W.~tau Yih, T.~Rocktäschel, S.~Riedel, and D.~Kiela, ``Retrieval-augmented generation for knowledge-intensive nlp tasks.'' arXiv:2005.11401 [cs.CL], 2021.

\bibitem{openai2023gpt4}
OpenAI, ``Gpt-4 technical report.'' arXiv:2303.08774 [cs.CL], 2023.

\bibitem{touvron2023llama}
H.~Touvron, L.~Martin, K.~Stone, P.~Albert, A.~Almahairi, Y.~Babaei, N.~Bashlykov, S.~Batra, P.~Bhargava, S.~Bhosale, D.~Bikel, L.~Blecher, C.~C. Ferrer, M.~Chen, G.~Cucurull, D.~Esiobu, J.~Fernandes, J.~Fu, W.~Fu, B.~Fuller, C.~Gao, V.~Goswami, N.~Goyal, A.~Hartshorn, S.~Hosseini, R.~Hou, H.~Inan, M.~Kardas, V.~Kerkez, M.~Khabsa, I.~Kloumann, A.~Korenev, P.~S. Koura, M.-A. Lachaux, T.~Lavril, J.~Lee, D.~Liskovich, Y.~Lu, Y.~Mao, X.~Martinet, T.~Mihaylov, P.~Mishra, I.~Molybog, Y.~Nie, A.~Poulton, J.~Reizenstein, R.~Rungta, K.~Saladi, A.~Schelten, R.~Silva, E.~M. Smith, R.~Subramanian, X.~E. Tan, B.~Tang, R.~Taylor, A.~Williams, J.~X. Kuan, P.~Xu, Z.~Yan, I.~Zarov, Y.~Zhang, A.~Fan, M.~Kambadur, S.~Narang, A.~Rodriguez, R.~Stojnic, S.~Edunov, and T.~Scialom, ``Llama 2: Open foundation and fine-tuned chat models.'' arXiv:2307.09288 [cs.CL], 2023.

\bibitem{jiang2024mixtral}
A.~Q. Jiang, A.~Sablayrolles, A.~Roux, A.~Mensch, B.~Savary, C.~Bamford, D.~S. Chaplot, D.~de~las Casas, E.~B. Hanna, F.~Bressand, G.~Lengyel, G.~Bour, G.~Lample, L.~R. Lavaud, L.~Saulnier, M.-A. Lachaux, P.~Stock, S.~Subramanian, S.~Yang, S.~Antoniak, T.~L. Scao, T.~Gervet, T.~Lavril, T.~Wang, T.~Lacroix, and W.~E. Sayed, ``Mixtral of experts.'' arXiv:2401.04088 [cs.LG], 2024.

\bibitem{nexusflow2023}
``Nexusraven-v2: Surpassing gpt-4 for zero-shot function calling.'' \url{https://nexusflow.ai/blogs/ravenv2}, December 2023.
\newblock Accessed on December 19, 2023.

\bibitem{userstories:2004}
M.~Cohn, {\em User Stories Applied: For Agile Software Development}.
\newblock USA: Addison Wesley Longman Publishing Co., Inc., 2004.

\bibitem{sommerville}
I.~Sommerville, {\em Software Engineering}.
\newblock Pearson, 10th~ed., 2015.

\bibitem{wppf}
W.~d.~P. Paula~Filho, {\em Software Engineering: Fundamentals, Methods and Standards – Projects and Processes (in Portuguese)}.
\newblock Gen-LTC Editora, 4th~ed., 2019.

\bibitem{voorhees2021guide}
D.~Voorhees, {\em Guide to Efficient Software Design: An MVC Approach to Concepts, Structures, and Models}.
\newblock Texts in Computer Science, Springer International Publishing, 2021.

\bibitem{nvidia2023}
T.~Varshney, ``Introduction to llm agents.'' \url{https://developer.nvidia.com/blog/introduction-to-llm-agents/}, November 2023.
\newblock Accessed on December 19, 2023.

\end{thebibliography}
\bibliographystyle{ieeetr}

\begin{table}
\vspace*{4pt}
\label{fig:raw-list}
\tiny
    \centering
    \begin{tabular}{ll}
    \hline
    Task & Instruction \\ \hline \\[1pt]
    Algorithm &
    Algorithm to check the availability of the selected pizza type in real-time \\[2pt]

    Algorithm &
    Algorithm to record the new order with a gourmet margherita pizza and a set time of \\ & 
    20 minutes from the current time \\[2pt]

    Algorithm &
    Algorithm to manage the countdown and ensure the order is ready in twenty minutes \\[2pt]

    Algorithm &
    Algorithm to notify the user when the order is placed, when it starts being prepared, \\ &
    and when it's ready for delivery or pickup \\[2pt]

    Algorithm &
    Algorithm to handle payment for the order through the app's integrated \\ & payment system \\[2pt]

    Algorithm &
    Algorithm to ensure the order is completed and pizza is handed off for delivery or \\ & pickup after 
    twenty minutes \\[2pt]

    Algorithm &
    Algorithm to filter pizzerias that offer gourmet Margherita pizzas \\[2pt]

    Algorithm &
    Algorithm to estimate delivery time based on user location and pizzeria location \\[2pt]

    Algorithm &
    Algorithm to filter pizzerias with an estimated delivery time of 20 minutes or less \\[2pt]

    Algorithm &
    Algorithm to check for promotions or discounts on a specific item \\[2pt]

    Algorithm &
    Algorithm to determine if quick delivery is available for an item \\[2pt]

    Algorithm &
    Algorithm that combines CheckPromotionForItem and ShowPromotionDetails for a \\ & specific item \\[2pt]

    Algorithm &
    Algorithm that combines CheckQuickDeliveryOption and ShowDeliveryOption for a \\ & specific item \\[2pt]

    Algorithm &
    Filter the customizations applicable to Margherita pizza \\[2pt]

    Algorithm &
    Filter customizations ensuring a 20-minute delivery \\[2pt]

    Algorithm &
    Algorithm that retrieves restaurants sorted by user ratings and filters for \\ & gourmet Margherita 
    pizza. \\[2pt]

    Algorithm &
    Algorithm that retrieves restaurant with the fastest delivery speed for \\ & Margherita pizza. \\[2pt]

    Algorithm &
    Algorithm that recommends the top-rated restaurant for gourmet Margherita \\ & pizza with 
    the fastest delivery. \\[2pt]

    Algorithm &
    Check availability of Margherita gourmet pizza \\[2pt]

    Algorithm &
    Calculate total cost for a single Margherita gourmet pizza including additional fees \\[2pt]

    Algorithm &
    Provide delivery time estimate for quick delivery option \\[2pt]

    Algorithm &
    Algorithm to filter restaurant data based on certain criteria \\[2pt]

    \hline \\[1pt]
    Data Source &
    Database table containing different types of pizzas including gourmet margherita \\[2pt]

    Data Source &
    Database table to store information about user orders including details and timings \\[2pt]

    Data Source &
    Model containing pizzeria information including location and menu offerings \\[2pt]

    Data Source &
    Data source representing promotions or discounts \\[2pt]

    Data Source &
    Data source representing menu items including pizzas \\[2pt]

    Data Source &
    Retrieve list of gourmet pizza customizations \\[2pt]

    Data Source &
    Retrieve delivery times for each customization option \\[2pt]

    Data Source &
    Data source containing restaurant details including ratings and reviews. \\[2pt]

    Data Source &
    Data source containing delivery speed information for restaurants. \\[2pt]

    Data Source &
    Retrieve minimum order requirements and additional fees \\[2pt]

    Data Source &
    Retrieve delivery options, time estimates, and fees for quick delivery \\[2pt]

    \hline \\[1pt]
    User Interface &
    Interface to display the PizzaMenu for user selection \\[2pt]

    User Interface &
    Interface to show confirmation details and allow users to confirm their order \\[2pt]

    User Interface &
    Interface to display the real-time status of the order including the countdown and \\ &
    readiness status \\[2pt]

    User Interface &
    Interface to display the list of nearby pizzerias that meet the criteria \\[2pt]

    User Interface &
    Interface to display promotion details to the user \\[2pt]

    User Interface &
    Interface to display quick delivery availability to the user \\[2pt]

    User Interface &
    Show available crust types and cheese options for Margherita pizza within 20-minute \\ &
    delivery time \\[2pt]

    User Interface &
    Interface to show the recommended restaurant to the user. \\[2pt]

    User Interface &
    Show availability, total cost, and delivery time for a single Margherita gourmet pizza \\[2pt]

    User Interface &
    Interface to show filtered restaurant results to the user \\[2pt]

    \end{tabular}
    \caption{Raw list of instructions created by an LLM}

\end{table}

\end{document}